\documentclass[onecolumn,useAMS,usenatbib]{mn2e}
\usepackage{amssymb}

\newcommand{\cs}{c_{\rm s}}

\newcommand{\kes}{\kappa_{\rm e.s.}}
\newcommand{\krel}{\hat\kappa}
\newcommand{\kB}{k_{\rm B}}
\newcommand{\Ledd}{L_{\rm E}}
\newcommand{\lum}{l_{\scriptscriptstyle\rm E}}
\newcommand{\Mach}{\mathcal{M}}
\newcommand{\dotsm}{\dot M_\odot}

\newcommand{\rpc}{r_{\rm pc}}
\newcommand{\rn}{R_{\rm N}}
\newcommand{\rs}{R_{\rm S}}
\newcommand{\rsg}{r_{\rm sg}}
\newcommand{\sigT}{\sigma_{\rm T}}
\newcommand{\Teff}{T_{\rm eff}}
\newcommand{\tth}{t_{\rm th}}
\newcommand{\tff}{t_{\rm ff}}
\newcommand{\veps}{\varepsilon}

\newcommand{\cm}{\,{\rm cm}}
\newcommand{\g}{\,{\rm g}}
\newcommand{\K}{\,{\rm K}}

\newcommand{\kms}{\,{\rm km~s^{-1}}}
\newcommand{\kpc}{\,{\rm kpc}}
\newcommand{\pc}{\,{\rm pc}}
\newcommand{\s}{\,{\rm s}}
\newcommand{\yr}{\,{\rm yr}}

\newcommand{\bc}{\begin{center}}
\newcommand{\ec}{\end{center}}
\newcommand{\bd}{\begin{displaymath}}
\newcommand{\ed}{\end{displaymath}}
\newcommand{\be}{\begin{equation}}
\newcommand{\ee}{\end{equation}}
\newcommand{\ba}{\begin{eqnarray}}
\newcommand{\ea}{\end{eqnarray}}
\newcommand{\bas}{\begin{eqnarray*}}
\newcommand{\eas}{\end{eqnarray*}}

\title{Selfgravity and QSO disks}
\author[Jeremy Goodman]
       {Jeremy Goodman \\
	Princeton University Observatory, Princeton, NJ 08544, USA}
\date{Received ??}

\pagerange{\pageref{firstpage}--\pageref{lastpage}}
\pubyear{2002}

\begin{document}

\maketitle

\label{firstpage}

\begin{abstract}
It is well known that the outer parts of QSO accretion disks are prone
to selfgravity if heated solely by orbital dissipation.  Such disks
might be expected to form stars rather than accrete onto the black
hole.  The arguments leading to this conclusion are reviewed.
Conversion of a part of the gas into high-mass stars or stellar-mass
black holes, and the release of energy in these objects by fusion or
accretion, may help to stabilize the remaining gas.  If the disk
extends beyond a parsec, however, more energy is probably required for
stability than is available by turning half the gas into high-mass
stars.  Small black holes are perhaps marginally viable energy
sources, with important implications (not pursued here) for the QSO
spectral energy distribution, the metallicity of the gas, microlensing
of QSO disks, and perhaps gravitational-wave searches.  Other possible
palliatives for selfgravity include accretion driven
by nonviscous torques that allow near-sonic
accretion speeds and hence lower surface densities for a given mass
accretion rate.  All such modes of accretion face major theoretical
difficulties, and in any case merely postpone selfgravity.
Alternatively, thin disks may not exist beyond a
thousand Schwarzshild radii or so (0.01 parsec), in which case QSOs
must be fueled by gas with small specific angular momentum.
\end{abstract}

\begin{keywords}
accretion disks---gravitation---quasars: general
\end{keywords}

\section{Introduction}

The outer parts of
steady, geometrically thin,
optically thick, viscously-driven accretion disks
around QSOs are predicted to be selfgravitating: that is,
the Toomre stability parameter $Q$ [eq.~(\ref{Qdef})] is much less
than unity; equivalently, the midplane density is much greater
than the Roche value \citep{Shlosman_Begelman87}.
As shown in \S\ref{alphadisk} for a standard $\alpha$ disk,
$Q$ falls to unity at a distance $r\sim 10^{-2}\pc$ from a
central black hole of mass $10^8 M_\odot$ accreting at the Eddington rate.
The mass of the disk within this radius is much less than
that of the black hole and hence able to fuel QSO activity for
only a fraction of the black-hole growth time, 
$t_{\rm Edd}\approx 5\times 10^7\veps_{0.1}^{-1}\yr$.
This is one reason to suspect that QSO disks are larger than
$10^{-2}\pc$, but there is also indirect observational evidence.
The spectral energy distribution of typical QSOs shows a
strong peak in the rest-frame infrared, which is often interpreted
as reprocessing of light from the central source by an extended
warped disk \citep{Sanders_etal89}.
Nearby Seyferts and other active galaxies, thought to be low-luminosity
analogs of QSOs, sometimes have resolvable nuclear disks.
Most spectacularly, VLBI observations of maser emission in NGC 4258
and NGC 1068 indicate disks on parsec scales and have been used to
measure the mass of their central black holes \citep{Nakai_Inoue_Miyoshi93,
Greenhill_etal95,Greenhill_Gwinn97}.

At least on larger galactic scales, it is believed that strong
self-gravity leads to star formation \citep{Martin_Kennicutt01}.
If that is true of QSO disks,
there is a danger that most of the gas would form stars, leaving little
to fuel the QSO.  Several theoretical attempts have been made to 
modify the standard $\alpha$-disk model so as to extend gaseous
QSO disks self-consistently into the selfgravitating regime.
The options include driving accretion with global bars or density
waves so as to increase the radial velocity and lower the surface
density needed to sustain a given mass accretion rate 
\citep{Shlosman_Begelman89}; heating or mechanically stirring the
disk with embedded stars or black holes so as to raise the
temperature and lower the density of the gas \citep{Collin_Zahn99a,
Collin_Zahn99b}; or allowing the disk to fragment into colliding
gas clumps whose epicyclic motions stabilize it against selfgravity,
at least on scales larger than the clumps themselves \citep{Kumar99}.

Our purpose is to re-examine selfgravity in QSO accretion disks
with an emphasis on dynamical and energetic constraints.
These constraints are most severe for massive and luminous systems,
so our interest is in black-hole masses $M\gtrsim 10^8 M_\odot$ and
luminosities close to the Eddington limit.

The plan of the paper is as follows.  \S2 reviews selfgravity in
steady $\alpha$ disks, including irradiation from the central source
(\S\ref{irradiation}), and enhancements to $\alpha$ by local
gravitationally-driven turbulence in a moderately selfgravitating disk
(\S\ref{gravturb}).  \S3 considers disks that are stabilized by
additional heating beyond that due to the dissipation of orbital
energy, but in which angular momentum continues to be transported by
an $\alpha$ viscosity.  The effective temperature of such disks falls
more slowly than the standard relation $\Teff\propto r^{-3/4}$; in
contrast to the usual situation, the total energy released per unit
mass accreted is strongly dependent on the outer radius of the disk
(\ref{Qdisk}).  \S4 briefly explores various alternatives to viscous
thin-disk accretion.  Schemes that involve thin disk but invoke
faster-than-viscous angular-momentum transport include global spiral
waves (\S\ref{spirals}) and magnetized disk winds (\S\ref{windy}).  In
every such case, we argue that the accretion velocity is bounded by
the sound speed.  Other alternatives explored in \S4 are less like
thin disks: quasi-spherical flows (\S\ref{hot}),
collisional star clusters (\S\ref{cluster}) and clumpy disks (\S\ref{clumpy}).

All of these alternatives to the standard alpha disk face severe
theoretical difficulties, or else seem unlikely to permit a centrifugally
supported accretion flow beyond $\sim 0.1\pc$.
We conclude in \S5  that the gas probably does not circularize beyond
this radius and must be supplied to the nucleus with low specific
angular momentum.

\section{Steady alpha disks}\label{alphadisk}

For the time being, we assume steady accretion and neglect
winds, so that the mass accretion rate is constant with radius,
and advection of angular momentum is balanced by viscous torque.

1. The mass accretion rate and viscosity parameter are related by
\be\label{alphadef}
\dot M=3\upi\alpha\beta^b\cs^2\Omega^{-1}\Sigma,
\ee
in which $\cs=\sqrt{p/\rho}$ 
is the isothermal sound speed at the disk midplane, $\Omega=(GM/r^3)^{1/2}$
is the orbital angular velocity, $\Sigma$ is the surface mass density,
and $\beta\equiv p_{\rm gas}/p$ is the ratio of gas pressure to total
pressure at the midplane.
The dimensionless \citet{Shakura_Sunyaev73} viscosity parameter $\alpha$ is
not necessarily constant but presumably $\lesssim 1$.  The
mechanism likely responsible for the viscosity of most disks
is magnetorotationally-driven turbulence,
for which simulations indicate $\alpha=10^{-3}-10^{-1}$
\citep{Balbus_Hawley98}.  As discussed below, where the disk
is selfgravitating, $\alpha$ may be as large as $\sim 0.3$.
Therefore, we often write $\alpha_{0.01}\equiv 10^{2}\alpha$
or $\alpha_{0.3}\equiv \alpha/0.3$.
The parameter ${b}$ is a switch that determines whether the
viscosity is proportional to gas pressure ($b=1$) or total pressure
($b=0$).  It is well known that radiation-pressure-dominated regions
of $\alpha$ disks have viscous instabilities in the latter case
\citep{Lightman_Eardley74}, but it is possible that $b=0$ in an
average sense.  Although the average surface density $\bar\Sigma$ would be
lower and the corresponding $\bar Q$ higher for $b=0$ than for $b=1$,
the viscous instability is expected to produce overdense
rings in which these trends may well be reversed.

2. If the disk were heated by viscous dissipation only,
then its surface temperature would be
\ba\label{Teff}
\Teff&=& \left(\frac{3}{8\upi\sigma}\frac{GM\dot M}{r^3}~\right)^{1/4}\\
&\approx& 2.9\times 10^3\, (M_8\dotsm)^{1/4}(r/10^{-2}\pc)^{-3/4}~\K,
\nonumber\\
&\approx& 6.2\times 10^5\, \left(\frac{\lum}{\veps_{0.1} M_8}\right)^{1/4}
\left(\frac{r}{\rs}\right)^{-3/4}~\K.\nonumber
\ea
We have introduced the abbreviations
$M_8\equiv M/(10^8 M_\odot)$, 
$\dotsm\equiv \dot M/(1\,M_\odot\,\yr^{-1})$,
dimensionless luminosity $\lum\equiv L/L_{\rm E}$,
radiative efficiency $\veps\equiv L/\dot Mc^2 \equiv 0.1\veps_{0.1}$,
and Schwarzshild radius $\rs=2GM/c^2\approx 10^{-5}M_8\,\pc$.
The mass accretion rate can then be written as
\be\label{mdotE}
\dot M= \frac{4\upi GM}{{\kes} c}\,\frac{\lum}{\veps}
~\approx 2.2\,\veps_{0.1}^{-1}\,\lum\, M_8\,\dotsm,
\ee
where ${\kes}\approx 0.4\cm^2\g^{-1}$ is the
electron-scattering opacity.

3. If viscous dissipation occurs mostly near the midplane, and vertical
transport of heat is by radiative diffusion, then the midplane
and surface temperatures are related approximately by
\be\label{Tmid0}
T^4\approx \frac{\kappa\Sigma}{2}\Teff^4.
\ee
Simulations of magnetorotational turbulence indicate that the
vertical scale height of the turbulent
dissipation is larger than that of the gas density \citep{Miller_Stone00}.
Hence eq.~(\ref{Tmid0}) may somewhat overestimate the midplane
temperature, in which case selfgravity may extend to
even smaller radii than estimated below.

\subsection{Onset of selfgravity}\label{sec_selfgravity}

The Toomre stability parameter for keplerian rotation is
\begin{equation}\label{Qdef}
Q=\frac{\cs\Omega}{\upi G\Sigma}\approx \frac{\Omega^2}{2\upi G\rho}\,,
\end{equation}
and local gravitational instability occurs where $Q<1$.
We have taken $\Sigma=2h\rho$ with
$h=c_s/\Omega$, which is approximately the disk half-thickness
when $Q\gtrsim1$.
Combining eq.~(\ref{Qdef}) with eq.~(\ref{alphadef}), one obtains
the simple relation
\be\label{MQrel}
G\dot M Q= 3\alpha\beta^{{b}}\cs^{3}\,.
\ee
For future reference, note that in a flat rotation curve, 
$v_{\rm circ}=r\Omega=\mbox{constant}$, the numerical factor 3 is
replaced by $2\sqrt{2}$ : a distinction too fine to matter.

To see why selfgravity is inevitable, consider the
radiation-pressure-dominated case, $\beta\ll 1$, so that
the isothermal sound speed $c_s^2=4\sigma T^4/3c\rho$.
From eqs.~(\ref{Teff}) \& (\ref{Tmid0}),
$T^4=3\kappa\Sigma\Omega^2\dot M/16\upi\sigma$.
Eliminating $T$ between these relations and 
using $\Sigma=2c_s\rho/\Omega$ leads to 
\begin{equation}\label{csrad}
c_s=\frac{\kappa\Omega\dot M}{2\upi c}
=\frac{\lum\kappa}{\veps\kes}\,\Omega\rs\qquad(\beta\ll 1),
\end{equation}
whence $h=(\lum/\veps)\rs=$constant for $\kappa=\kes$.
Using this for $c_s$ and eq.~(\ref{mdotE}) for $\dot M$
in eq.~(\ref{MQrel}) leads to
\begin{equation}\label{Qrad0}
Q= \frac{3\alpha\beta^b}{8\upi\sqrt{2}}\left(\frac{\lum}{\veps}\right)^2
\krel^3\left(
\frac{{\kes}c^4}{G^2M}\right)\left(\frac{\rs}{r}\right)^{9/2},
\qquad(\beta\ll 1),
\end{equation}
in which $\krel\equiv\kappa/\kes$.
Hence disks around more massive black holes are more prone to selfgravity.
For $b=0$, there is a characteristic mass 
above which an Eddington-limited disk would be selfgravitating even at
its inner edge.
This mass is enormous ($\sim 10^{19} M_\odot$ for $\alpha_{0.01}
=\veps_{0.1}=1$), but even for realistic black-hole masses, $Q<1$ at
radii greater than 
\begin{equation}\label{rsgrad}
r_{\rm s.g.} \approx 2.1\times 10^3\,\left(\frac{\alpha_{0.3}\beta^b\lum^2
\krel^3}{\veps_{0.1}^2 M_8}\right)^{2/9}~\rs\qquad(\beta\ll1).
\end{equation}
Notice that $\alpha$ has been scaled to $0.3$ rather than $10^{-2}$
because of the expected enhancement by gravitational turbulence (see
below).

We now examine the assumption that $\beta\ll1$.  It follows from
eqs.~(\ref{mdotE}), (\ref{Qdef}), and (\ref{MQrel}) that
\begin{eqnarray}\label{betaQ}
\frac{\beta^{1-(b/2)}}{(1-\beta)^{1/4}}&=& 2^{-5/2}
\left(\frac{3}{\upi}\right)^{3/4}\left(\frac{\veps\alpha}{\lum}\right)^{1/2}
Q^{-3/4}\left(\frac{\kB^4c^9{\kes}^2}{m^4G^7\sigma}\right)^{1/4}
\left(\frac{r}{\rs}\right)^{-3/4}\,M^{-1}\\
&\approx& 0.39\left(\frac{\veps_{0.1}\alpha_{0.3}}{\lum}\right)^{1/2}
Q^{-3/4}\left(\frac{r}{2100\rs}\right)^{-3/4}\,M_8^{-1}\,.\nonumber
\end{eqnarray}
Eq.~(\ref{betaQ}) does not depend upon
eqs.~(\ref{Teff}) \& (\ref{Tmid0}): that is, it does not assume that
radiative losses are balance by viscous dissipation.  (The opacity
$\kes$ enters the formula only because it is used in the definition
of the Eddington ratio $\lum$).
Hence eq.~(\ref{betaQ}) remains true even when auxiliary sources
of heating are present in the disk.  For viscously heated, radiatively
cooled disks, $\beta$ is given by eq.~(\ref{betaval}) in the Appendix.

\subsection{Irradiation}\label{irradiation}

Although flared or warped outer regions can
be warmed somewhat by irradiation from the inner parts (where
most of the total disk luminosity originates), we now show
this effect is not enough to stabilize the disk at $r\sim r_{\rm s.g.}$.

To do so, irradiation must raise the midplane
temperature $T$, not just the surface temperature $\Teff$, substantially.
As a result, the disk will be nearly isothermal from surface to midplane.
The vertical pressure \emph{gradient}, and therefore the disk thickness,
will be dominated by the gas even if $\beta\ll 1$: that is, $h^2\approx 
p_{\rm gas}/\rho\Omega^2$.
If the effective viscosity derives from local magnetorotational
turbulence, then it probably scales more directly with
thickness than with sound speed since MRI instabilities do not require
compression.  Then in all relevant respects
the disk behaves as if $\beta=1$ and $c_s^2\to p_{\rm gas}/\rho=\kB T/m$.
From eq.~(\ref{MQrel}), it then follows that the minimum temperature
that the irradiation must provide to ensure gravitational stability is
\begin{equation}\label{T11}
T_{Q=\beta=1}= \frac{m}{\kB}\left(\frac{G\dot M}{3\alpha}\right)^{2/3}
\approx 5.6\times 10^4\left(\frac{\lum M_8}
{\alpha_{0.3}\veps_{0.1}}\right)^{2/3}~\K.
\end{equation}
(To be conservative, we have
scaled to $\alpha=0.3$; for $\alpha=0.01$, this temperature would
be an order of magnitude larger).  On the other hand, the temperature
of the disk in equilibrium with radiation from the central source is
\begin{equation}\label{Teq}
T_{\rm eq}= \left(\frac{c^5\lum\cos\theta}{4\sigma\kes GM}\right)^{1/4}
\left(\frac{r}{\rs}\right)^{-1/2}\approx
3.8\times10^5 \left(\frac{\lum\cos\theta}{M_8}\right)^{1/4}
\left(\frac{r}{\rs}\right)^{-1/2}\,\K,
\end{equation}
where $\theta$ is the angle between the local normal to the
disk and the radial rays.  For a flat disk, 
$\cos\theta\sim\max(h,\rs)/r$ at $r\gg\rs$, but even
for a severely warped disk with
$\cos\theta\sim 1$, $T_{\rm eq}\ll T_{Q=\beta=1}$ at 
$r/\rs\gg 44(\alpha_{0.3}\veps_{0.1})^{4/3}\lum^{-5/6}M_8^{-11/6}$.
Since this last inequality certainly holds at all $r\ge r_{\rm s.g.}$,
we conclude that irradiation is not important for the selfgravity
of the disk.

\subsection{Local gravitational turbulence}\label{gravturb}

It has occasionally been proposed that a partially selfgravitating disk
may transport angular momentum by spiral waves
\citep[\emph{e.g.}][]{Cameron78,Paczynski78}.
Others have suggested that gravitational instabilities are intrinsically
global and therefore not reducible to a local viscosity 
\citep{Balbus_Papaloizou99}.  It may be that both opinions are
sometimes correct, depending upon the ratio of disk thickness
to radius, since the most unstable wavelength in a $Q=1$ disk
is $\sim h$.  We consider local gravitational turbulence here,
since it can be accommodated within the $\alpha$ model; a (much larger)
upper bound on angular momentum transport by global spirals is given
in \S\ref{spirals}. 

By careful two-dimensional simulations,
\cite{Gammie01} finds
that gravitationally-driven turbulence can be local
and can support $\alpha$ approaching unity; in fact, in the
absence of any other viscosity mechanism, his disks self-regulate
themselves so that 
\be\label{alphagrav}
\alpha_{\rm grav}\approx \frac{1}{(\gamma_{\rm 2D}-1)\Omega\tth},
\ee
where $\tth\equiv\Sigma\kB T/\sigma\Teff^4$ is the local thermal timescale,
and $\gamma_{\rm 2D}$ is the 2D adiabatic index: 
\begin{displaymath}
\gamma_{\rm 2D}= \left.\frac{\partial}{\partial\log\Sigma}
\right)_{S}\log\left[\int\limits_{-\infty}^\infty p(r,\theta,z)\,dz\right].
\end{displaymath}
[Our eq.~(\ref{alphagrav}) differs
from Gammie by a factor $\gamma_{\rm 2D}$ because we define
$\alpha$ in terms of isothermal rather than adiabatic sound speed.]
Gammie finds that self-regulation fails and the disk fragments
if $\Omega\tth\lesssim 0.3$ for $\gamma_{\rm 2D}=2$.
The latter result is not entirely surprising: in the absence of
pressure support, the natural timescale for
collapse is $t_{\rm dyn}=\Omega^{-1}$, and one would not expect
collapse to be prevented by thermal energy unless $\tth>t_{\rm dyn}$.
The converse statement, that fragmentation can be indefinitely postponed if
$\Omega\tth>0.3$, is not at all obvious but is supported by
Gammie's simulations.

We are not aware of any direct numerical simulations of disks that
are unstable to both gravitational and magnetorotational
modes at the same radii,
as is likely to be the case for QSO disks \citep{Menou_Quataert01}.
It would be interesting to explore this, as there might be a
synergy between the two instabilities.

QSO disks are expected to be very thin in the regions
of interest to us, so that Gammie's results may be applicable.
Eq.~(\ref{csrad}) implies $h=\lum\rs/6\veps$ in
a radiation-pressure dominated disk, and hence 
$h/r\lesssim 0.003$ at the innermost radius where $Q\sim 1$ 
[eq.~(\ref{rsgrad})].
We expect that from this radius outwards, $\alpha$ will rise smoothly
from the value supported by magnetorotational turbulence (perhaps
$\alpha_{\rm m.h.d.}\sim 10^{-2}$) up to the maximum allowed by
gravitational turbulence, $\alpha_{\rm grav,max}\sim 0.3$, in
such a way that $Q\approx$constant.  It follows from eq.~(\ref{Qrad0})
that the ratio of outer to inner radii of this region is rather modest:
$\sim (\alpha_{\rm grav,max}/\alpha_{\rm m.h.d.})^{2/9}\sim 2$.
At still larger radii, additional sources of energy are required in
order to prevent catastrophic fragmentation.

\section{Constant-$Q$ disks}\label{Qdisk}

In this section, we postulate that some heating process arises that
maintains the disk on threshold of gravitational instability, \emph{i.e.}
at a constant $Q\approx 1$,
and work out some consequences for the structure of the disk.
In \S\ref{subsec_radialQ}, we discuss the
radial dependence of midplane temperature, density, and $\beta$ [already
given by eq.~(\ref{betaQ})].
In \S\ref{energetics}, we estimate the minimal amount of energy in excess
of viscous dissipation that must be supplied to maintain constant $Q$.

For our immediate purposes, the details of the heating process are not
important as long as it provides a stable feedback that regulates $Q$.
Angular momentum is assumed still to be viscous,
though perhaps with an enhanced $\alpha$ in eq.~(\ref{alphadef}).
The effective temperature of the disk may no longer be determined by equation
(\ref{Teff}), however, because viscous dissipation may not dominate
the thermal energy budget.  

Although they may seem artificial, these assumptions are probably
appropriate for the disks of spiral galaxies, and for the
local interstellar medium in particular.  The local
Galactic magnetic field is consistent with simulations
of magnetorotational turbulence: {\it viz.}
a somewhat sub-equipartition strength, a predominantly toroidal
orientation, and fluctuations comparable to the mean
\citep{Brandenburg_etal95}.  It is plausible therefore
that there is a nonzero average magnetic stress 
$\langle B_r B_\theta\rangle/4\upi=\alpha p_{\rm gas}$
that systematically
transfers angular momentum outwards \citep{Sellwood_Balbus99}.
Taking $ p_{\rm gas}/\kB\approx 2000\,\cm^{-3}\,\K^{-1}$,
$\rho\approx 0.3\,m_H\,\cm^{-3}$ 
\citep{Spitzer78}, and circular velocity $V_0\approx 200\,\kms$
\citep{Binney_Merrifield98}, the implied radial drift velocity
$v_r\approx -\alpha p_{\rm gas}/\rho V_0\approx -0.3\,\alpha\,\kms$ is
small enough to have escaped detection.
Perhaps coincidentally, 
eq.~(\ref{MQrel}) predicts 
$\dot M\approx 3\times 10^{-2}\alpha_{0.1}Q^{-1}\dot M_\odot\yr^{-1}$,
about half the Eddington rate for the Galaxy's $2.5\times 10^6 M_\odot$
central black hole.
The implied viscous heating rate 
$\alpha p_{\rm gas}\Omega_0\approx 2\times 10^{-29}\,\alpha_{0.1}\,
\mathrm{erg\,cm^{-3}}$ is negligible compared to the inferred
radiative cooling rate of the gas, $\sim 2\times 10^{-26}
\mathrm{erg\,cm^{-3}}$ \citep{Spitzer78}.  Presumably,
the temperature of the ISM is maintained by stars.  An important
difference between the the local ISM and QSO disks is that the
former is very optically thin, especially to absorption, which means
that the energy input from stars is inefficiently radiated.

\subsection{Density and temperature}\label{subsec_radialQ}

The midplane density in a constant-$Q$ disk follows from
eq.~(\ref{Qdef}):
\begin{equation}\label{rhoQ}
\rho= \frac{M}{2\upi Q r^3} = 1.2 M_8^{-2}Q^{-1}\left(\frac{\rs}{r}\right)^{3}
~\mathrm{g\,cm^{-3}}\,,
\end{equation}
so that the density at $r_{\rm s.g.}$ [eq.~(\ref{rsgrad})] is
$\sim 10^{-8} M_8^{-4/3}\,\mathrm{g\,cm^{-3}}$.
The ratio $\beta/(1-\beta)= 4\sigma T^3/3c\rho$ is determined by
eq.~(\ref{betaQ}).
The temperature itself is (for $\beta\ll 1$)
\begin{eqnarray}\label{TQ0}
T&=&
2^{-7/6}\left(\frac{3}{\upi}\right)^{1/12}\left(\frac{\lum}{\alpha
\veps Q^{1/2}}\right)^{1/6}c^{19/12}G^{-5/12}\sigma^{-1/4}{\kes}^{-1/6}
\,M^{-1/3}\left(\frac{\rs}{r}\right)^{3/4}\nonumber\\
&\approx& 6.9\times 10^6 \left(\frac{\lum}{\alpha_{0.3}\veps_{0.1}Q^{1/2}}
\right)^{1/6}M_8^{-1/3}\left(\frac{\rs}{r}\right)^{3/4}\,\K
\qquad\mathrm{if}~b=0\,;\\[5ex]
\label{TQ1}
T&=&2^{-1/3}\left(\frac{\upi}{3}\right)^{1/6}\left(\frac{\lum Q^{1/2}}{\alpha
\veps}\right)^{1/3}c^{5/6}G^{1/6}\sigma^{-1/6}{\kes}^{-1/3}(\kB/m)^{-1/3}
\,\left(\frac{\rs}{r}\right)^{1/2}\nonumber\\
&\approx& 1.4\times 10^6 \left(\frac{\lum Q^{1/2}}{\alpha_{0.3}\veps_{0.1}}
\right)^{1/3}\left(\frac{\rs}{r}\right)^{1/2}\,\K
\qquad\mathrm{if}~b=1\,.
\end{eqnarray}
The surface density is
\begin{eqnarray}\label{SigmaQ0}
\Sigma&=&\left(\frac{\lum\, c^8}{12\pi^2\sqrt{2}\,\veps\alpha Q^2\,
G^4\kes M^2}\right)^{1/3}\left(\frac{\rs}{r}\right)^{3/2}\nonumber\\
&\approx& 7.4\times 10^9\left(\frac{\lum}{\alpha_{0.3}\veps_{0.1}Q^2}
\right)^{1/3} M_8^{-2/3}\left(\frac{\rs}{r}\right)^{3/2}\,
\mbox{g cm}^{-2}\qquad\mbox{if}~b=0;
\end{eqnarray}
\begin{eqnarray}\label{SigmaQ1}
\Sigma&=&\frac{2}{3}\left(\frac{2\lum\,m}{\veps\alpha\,\kB\kes}\right)^{2/3}
\left(\frac{3\,\sigma c^7}{\upi Q\,G}\right)^{1/6}
\left(\frac{\rs}{r}\right)\nonumber\\
&\approx& 5.4\times 10^8\,Q^{-1/6}\left(\frac{\lum}{\alpha_{0.3}\veps_{0.1}}
\right)^{2/3} \left(\frac{\rs}{r}\right)\,
\mbox{g cm}^{-2}\qquad\mbox{if}~b=1.
\end{eqnarray}
For the nominal values of the parameters shown, 
$\Sigma$ would fall to $1\,\mbox{g cm}^{-2}$ at $40\pc$ for $b=0$
and $5\kpc$ for $b=1$; however, in the latter case $h/r$ rises
more rapidly with $r$ and is already $\sim 0.5$ at $r=1\pc$.

Beyond $10^4-10^5\rs\approx 0.1-1\,M_8\,\mathrm{pc}$, the above formulae
predict $T\lesssim 5000\,\K$, so that the opacity $\kappa\ll\kes$ 
and the disk becomes optically
thin.  (This assumes $M_8=\lum=\veps_{0.1}=\alpha_{0.3}=1$.  Dust
will raise opacity again at $T\lesssim 1700\K$.)
The disk must then be supported by gas pressure,
notwithstanding eq.~(\ref{betaQ}) which presumes that the radiation is
trapped.
But at $\beta=1$, the minimum temperature for gravitational stability
is (\ref{T11}), which is about an order of magnitude larger than
predicted by the formulae above.
So in a marginally gravitationally stable disk,
there must be an extended region where the temperature
adjusts itself within a limited range
($5000\K\lesssim T\lesssim 10^4\K$)
so that the disk is marginally optically thin.
At the low densities relevant here, the maximum opacity is
$\kappa_{\rm max}\approx 10\kes$ and is achieved at $T\approx 10^4\K$
\citep{Kurucz92,Keady_Kilcrease00}.
Hence the outer edge of the region in question should end at
$\Sigma\approx \kappa_{\rm max}^{-1}\approx 0.3\,\mathrm{cm^2\,g^{-1}}$
which occurs (assuming $Q=1$ and $T=10^4\K$) at
\begin{equation}\label{rthin}
r_{\rm thin}\approx\frac{\cs\kappa_{\rm max}v_{\rm circ}}{\upi G}
\approx 170 M_8^{0.24}\,\pc,
\end{equation}
provided of course that the disk extends to such large radii.
We have taken the circular velocity as $\sigma_{\rm bulge}\sqrt{2}
\propto M^{0.24}$  [see eqs.~(\ref{Gebrel})\&(\ref{MFrel})] 
rather than $\sqrt{GM/r}$, since $r_{\rm thin}$
lies outside the black hole's sphere of influence
[eq.~(\ref{rn})].

The parts of the disk beyond $r_{\rm thin}$ would radiate predominantly
in optical emission lines with velocity widths $\sim\sigma_{\rm bulge}$,
so that they might be identified with the QSO narrow-line region.
However, we will now see that the energy required to maintain
the disk at constant $\dot M$ and $Q\gtrsim 1$ all the way out to
$r_{\rm thin}$ would be prohibitive.

\subsection{Energetics}\label{energetics}

We define a local disk efficiency by
\begin{equation}\label{epslocal}
\veps'(r)\equiv \frac{4\upi r^2\sigma\Teff^4}{3\dot M c^2}.
\end{equation}
For a viscously heated disk in newtonian gravity, $\veps'c^2$ reduces to the
local binding energy per unit mass, $GM/2r=c^2\rs/4r$ [see eq.~(\ref{Teff})],
which is largest at small radii; given a torque-free inner boundary at
$r=r_{\rm min}$, the global efficiency is $\veps=\veps'(r_{\rm min})$.
But the constant-$Q$ disks require additional energy inputs, so that
$\veps'$ is generally larger than $\rs/4r$ and tends in fact
to $\emph{increase}$ with radius.

Using the results of \S\ref{subsec_radialQ} for the midplane temperature
$T$, and assuming that vertical radiative transport obeys eq.~(\ref{Tmid0}),
we find that
\begin{eqnarray}\label{epsQ0}
\veps'(r) &=& \frac{\upi^{1/3}}{2^{7/6}3^{1/3}}\,\krel^{-1}\,
\left(\frac{Q\veps^2}{\alpha l^2}
\,\frac{G\rs}{c^2\kes}\right)^{1/3}\,\left(\frac{r}{\rs}\right)^{1/2}
\nonumber\\
&\approx& 8.2\times 10^{-4}\left(\frac{Q\veps_{0.1}^2}{\alpha_{0.3}\lum^2}
\right)^{1/3}\,\krel^{-1}M_8^{1/3}\left(\frac{r}{10^5\rs}\right)^{1/2}
\qquad\mathrm{if}~b=0;\\[5ex]
\label{epsQ1}
&=& \frac{\upi^{5/6}}{2^{1/6}3^{5/6}}\,\frac{\kes^{1/3}}{\kappa}\,
\,\frac{Q^{5/6}\veps^{1/3}}{\alpha^{2/3}\lum^{1/3}}
\left(\frac{G}{c}\right)^{5/6}\left(\frac{m}{\kB}\right)^{2/3}\sigma^{1/6}
\,r
\nonumber\\
&\approx& 1.1\times 10^{-2}\left(\frac{Q^{5/2}\veps_{0.1}}{\alpha_{0.3}^2\lum}
\right)^{1/3}\,\krel^{-1}M_8\left(\frac{r}{10^5\rs}\right)
\qquad~~~~\mathrm{if}~b=1.
\end{eqnarray}
It would appear from these relations that $\veps'\to\infty$ as
$\kappa\to 0$, but they are not valid when the disk is optically thin.
If the absorption optical depth $\tau=\kappa\Sigma <1$,
then $\Teff^4\approx \tau T^4$ rather than $T^4/\tau$.
So in fact $\veps'\to0$ as $\kappa\to0$.
Hence the largest radius at which eqs.~(\ref{epsQ0}) or (\ref{epsQ1})
is valid is the smallest among the radii $r_{\rm rthin},\rn$, and 
$r_{\rm out}$,
where the first is the radius at which a $Q=1$ disk would
become optically thin [eq.~(\ref{rthin})], the second is the radius
within which the black hole dominates the potential,
\begin{equation}\label{rn}
\rn\equiv \frac{GM}{2\sigma_{\rm bulge}^2}\approx 
(6\pm 1)\, M_8^{0.5\pm0.1}\,\pc\,,
\end{equation}
and the last is the actual outer edge of disk, which depends
upon the angular momentum of the gas supplied to it.
The final expression in eq.~(\ref{rn}) has used the recently reported
$M-\sigma_{\rm bulge}$ relation 
\citep[see \S\ref{cluster} below]{Gebhardt_etal00,Merritt_Ferrarese01}.
Actually, eq.~(\ref{rn}) probably underestimates the distance to which
the black hole dominates the circular velocity, because the density
profile in the cusp of bright bulges is considerably shallower than
$r^{-2}$.  

At $r=10\,\pc\approx 10^6\,M_8^{-1}\rs$, equations~(\ref{epsQ0}) 
and (\ref{epsQ1}) imply $\veps'\approx 0.003\,M_8^{-1/6}$ and
$\veps'\approx 0.1$, respectively.  The former efficiency is barely
compatible with thermonuclear burning even if all of the disk gas
is processed through high-mass stars.  The latter is unsustainable
by stars and in fact comparable to the efficiency of the central engine.

\section{Alternatives to thin disks}\label{alternatives}

As shown in \S\ref{alphadisk}, a geometrically thin, optically thick
accretion disk in a typical high-luminosity QSO would be
selfgravitating at radii $r\gtrsim 10^2-10^3\rs$, or
$10^{-3}-10^{-2}\pc$, where $\rs=2GM/c^2$ is the Schwarzschild radius
of the central black hole.  Large
global radiative efficiency, $\veps\gtrsim 0.1$, probably requires a thin
disk near $\rs$, but it does not much depend upon the nature of the
flow at large radius.
So in this section, we consider whether
the selfgravitating part of the disk can be replaced
by some other form of accretion.

\subsection{Hot, quasi-spherical flows}\label{hot}

These have been proposed as models for accretion at very low accretion
rate ($\dot M$) and low radiative efficiency $\veps\equiv L/\dot
Mc^2$
\citep{Rees_Phinney_Begelman_Blandford82,Ichimaru77,Narayan_Yi94}.  At
sufficiently low density, the gas cooling time and ion-electron
thermal equilibration time are longer than the accretion time, so that
the ion temperature ($T_i$) is approximately virial and the thickness
of the disk is comparable to its radius.  One supposes that
angular-momentum transport is efficient (viscosity parameter
$\alpha\sim 1$), or that the angular momentum is very small to begin
with, so that the accretion velocity ($v_r$) is comparable to the
free-fall velocity.  The large $v_r$ and $T_i$ combine with the low
$\dot M$ to make the density low, as required.  It is unclear just how
low $\dot M$ must be for this mode of accretion to sustain itself,
because angular-momentum transport is not well understood and
collisionless processes may enhance the thermal coupling of ions and
electrons.

A quasi-spherical flow is not viable when the
luminosity is close to the Eddington limit, however,
because of inverse-Compton cooling.
For spherical free fall onto a source of luminosity $L=\lum L_{\rm Edd}$,
the inverse-Compton cooling rate of free electrons is
\begin{equation}\label{tic}
t_{\rm C}^{-1} \approx \frac{2\sigT L}{3\upi r^2 m_e c^2}
=\frac{8}{9}\frac{m_p}{m_e}\lum\left(\frac{\rs}{r}\right)^{1/2}
\,\tff^{-1}\,.
\end{equation}
We have introduced the free-fall time of a radial parabolic
orbit from radius $r$,
\begin{equation}\label{tff}
\tff\equiv \frac{2}{3}\left(\frac{r^3}{2GM}\right)^{1/2}~
\approx 2.2\times 10^{10}\,M_8^{-1/2}\rpc^{3/2}\,\s,
\end{equation}
in which $\rpc$ is the distance from the source in parsecs and
$M_8\equiv M/(10^8 M_\odot)$.
Actually, radiation pressure increases the free-fall time by
$(1-\lum)^{-1/2}$, but this factor is neglected for simplicity.
From eq.~(\ref{tic}), $t_{\rm C}<\tff$ at
$r< 25\lum^2 M_8\pc$.
Hence the electrons assume
the color temperature of the central source (or even less, see below):
$T_{\rm C}\lesssim 10^5 M_8^{-1/4}$.  This is much less than the
virial temperature, $T_{\rm vir}\approx 10^7 M_8 \rpc^{-1}$, at
all radii of interest to the present paper.
\footnote{The timescale for Poynting-Robertson drag is
$\approx (m_p/m_e)t_{\rm C}$,
hence $\gg\tff$ at $r\gg\rs$, \emph{i.e.}, the radiation field does
not remove angular momentum from the gas fast enough to prevent
it from circularizing.}
The electron density is
\begin{equation}\label{n_e}
n_e=\left(\frac{\lum}{\veps}\right)\frac{1}{3c\sigT\,\tff},
\end{equation}
so that the electron-ion equilibration time due to Coulomb
collisions alone \citep{Spitzer78} is much shorter than the flow time:
\begin{displaymath}
\frac{t_{\rm eq}}{\tff}\approx 2\times 10^{-5}
\veps\lum\left(\frac{T_e}{10^5\K}\right)^{3/2}~.
\end{displaymath}
Of course $T_e\sim 10^5\K$ is the peak of the cooling curve
\citep{Spitzer78}.  If $n_e$ obeys eq.~(\ref{n_e}), then 
radiative cooling is actually faster than
inverse-Compton cooling:
\begin{displaymath}
\frac{t_{\rm rad}}{t_{\rm C}}\approx 6.\veps
\left(\frac{\rs}{r}\right)^{1/2}\qquad(T_e=10^5\K).
\end{displaymath}

In short, both the ions and the electrons of a quasi-spherical flow
onto a near-Eddington QSO cool in much less than a free-fall time.
A thin disk will form unless the specific angular momentum of the flow is
negligible.

\subsection{Collisional stellar cluster}\label{cluster}

Dense stellar clusters have occasionally been nominated as precursors
to QSO black holes, either by relativistic collapse
\citep{Zeldovich_Podurets66,Shapiro_Teukolsky85,Ebisuzaki_etal01}
or by collisions among non-degenerate stars \citep{Spitzer_Saslaw66,Rees78}.
Our interest, however, is in stellar collisions as the main source of fuel
for an already very massive black hole.  This has been studied by
\citet{McMillan_Lightman_Cohn81} and \citet{Illarionov_Romanova88},
among others, who show that in order to supply a $10^8 M_\odot$ 
black hole at its Eddington rate, the velocity dispersion of such
a cluster must be $\gtrsim 10^3\kms$.   To demonstrate the robustness
of this conclusion, we will make some oversimplified but conservative
estimates here.

In order to provide high radiative efficiency ($\veps$), stars must be
disrupted and their gaseous debris circularized before accretion.  If
$M_{\rm b.h.}\lesssim 10^8 M_\odot$, main-sequence stars scattered
onto loss-cone orbits are likely to be tidally disrupted rather than
swallowed whole \citep{Hills75}.  About half of the tidal debris is
unbound and promptly escapes from the black hole
\citep{Lacy_Townes_Hollenbach82,Evans_Kochanek89}, and much of the
remainder is likely to be swallowed at low radiative efficiency before
the gas circularizes \citep{Cannizzo_Lee_Goodman90,
Ayal_Livio_Piran00}.  We limit our discussion to $M_{\rm b.h.}\gtrsim
10^8 M_\odot$, as required for the most luminous QSOs, and assume that
stars are disrupted by stellar collisions.  We ignore loss-cone
effects because stars swallowed whole do not contribute to the QSO
luminosity.

The total collision rate among $N$ stars forming a cluster of
structural length $a$ is
\begin{equation}\label{Ndot}
\dot N = {C}\,\bar\sigma^7\,N^2(GM)^{-3}R_*^{2},
\end{equation}
where ${C}$ is a dimensionless coefficient,
$\bar\sigma$ the root-mean-square 
velocity dispersion in one dimension averaged over the cluster,
$N$ the number of stars, and $R_*\sim R_\odot$ the radius of an
individual star.
$M$ is the total mass that determines $\bar\sigma$ \emph{via}
the virial theorem, so
that if $m_*$ is the mass of individual stars and $M_*\equiv Nm_*$
then $M\approx M_{\rm bh}+\mbox{$1\over2$}M_*$, the
factor of $1/2$ being needed to avoid double-counting the gravitational
interactions among the stars.
We take $4\upi R_*^2$ for the collision cross section; this
allows for grazing collisions that probably would not disrupt the stars
\citep{Spitzer_Saslaw66},
but it will lead to a conservative estimate of $\bar\sigma$.
Gravitational focusing is unimportant at the high velocity
dispersions relevant here.

A second relation for $\dot N$ follows by requiring the
collisional debris to sustain the QSO at a fraction 
$\lum\equiv L/\Ledd$ of its Eddington luminosity:
\begin{equation}\label{colldebris}
2M_*\dot N = \frac{\lum}{\veps}\frac{4\upi GMm_p}{c\sigT}.
\end{equation}
Eliminating $\dot N$ between eqs.~(\ref{Ndot}) and (\ref{colldebris})
leads to
\begin{displaymath}
\bar\sigma^7= \frac{2\upi\lum}{{C}\veps}\,\frac{G^4m_p}{\sigT c}\,
\frac{m_*}{R_*^2}\left(\frac{M_{\rm b.h.}M^3}{M_*^2}\right)~.
\end{displaymath}
The term in parentheses achieves its minimum value $27 M_{\rm b.h.}^2/16$
at $M_*=4M_{\rm b.h.}$.

The coefficient ${C}$ depends upon the density profile $\rho(r)$ of the
stellar cluster.  It can be arbitrarily large if $\rho(r)$ is
sufficiently steep, but then the collision rate is dominated by
stars at small radius.
These tightly-bound stars represent only a small fraction of the cluster
mass and would be consumed in much less than the growth time
of the black hole, unless their total mass 
($M_{\rm t.b.}$) is $\gtrsim M_{\rm b.h.}$,
in which case the tightly-bound population is
substantially self-gravitating and we redefine $M_*\equiv M_{\rm t.b.}$.
One might suppose that $M_{\rm t.b.}\ll M_{\rm b.h.}$ and that
the tightly-bound stars were continuously
replenished by two-body relaxation from a reservoir of more weakly
bound stars; but the larger cluster would then 
have to expand to conserve energy, reducing
the collision rate and the fueling of the QSO.  For these reasons,
we consider clusters for which ${C}$ is dominated by stars near
the half-mass radius.
For example, if the stars have isotropically distributed
orbits with a common semimajor axis $a$ in a keplerian potential,
then $\rho(r)\propto\sqrt{(2a/r)-1}$, $\bar\sigma^2=GM/3a$, and
${C}\approx 196.$  This leads to
\begin{equation}\label{sigma_est}
\bar\sigma\approx 760\left(\lum\over 10\veps\right)^{1/7}
\left(g_*\over g_\odot\right)^{1/7}\, M_8^{2/7}\,\kms,
\end{equation}
where $g_*$ is the stellar surface gravity.  Because of the
one-seventh root, the result is not very sensitive to our assumptions.
A Plummer sphere of the same total mass, 
$M_*=5\times 10^8 M_\odot$, yields $\bar\sigma\approx 730\kms$.

Equation (\ref{sigma_est}) can be compared with recently-discovered
empirical relations between inactive black holes---presumably
QSO relics---and their host bulges.
\cite{Gebhardt_etal00} find
\begin{equation}\label{Gebrel}
M_{\rm b.h.}= 1.2(\pm0.2)\times 10^8
\left(\sigma_{\rm e}\over 200\kms\right)^{3.75\pm0.3}\,M_\odot\,,
\end{equation}
where $\sigma_{\rm e}$ is the line-of-sight velocity dispersion at
one effective radius.
\cite{Merritt_Ferrarese01} use the central velocity dispersion,
which is usually little different from $\sigma_{\rm e}$:
\begin{equation}\label{MFrel}
M_{\rm b.h.}= 1.30(\pm0.36)\times 10^8
\left(\sigma_{\rm c}\over 200\kms\right)^{4.72\pm0.36}\,M_\odot\,.
\end{equation}
The scaling exponent $d\log M/d\log\sigma=3.5$
implied by eq.~(\ref{sigma_est}) is similar to the empirical ones
(\ref{Gebrel}) \& (\ref{MFrel}), but the normalization is very
different.  Extrapolated to $700\kms$, the empirical relations
predict $M_{\rm b.h.}\gtrsim 10^{10}M_\odot$ instead of $10^8 M_\odot$.

Therefore, if QSOs were fueled by dense stellar clusters, these
clusters must have been an order of magnitude more tightly bound than
the surrounding bulge, and they would have been a dynamically distinct
stellar component.  There seems to be very little trace of this
tightly-bound stellar population in present-day bulges.  How would
such a component form?  A likely possibility is gaseous dissipation
followed by star formation.  As will be seen, accretion in a thin
viscous disk may lead to just such a result.  We have discussed
fueling the QSO by a disk and fueling it by stellar collisions as
though these were mutually exclusive possibilities, but perhaps the
two occur in concert.

\subsection{Wind-driven disks}\label{windy}

In principle at least, a magnetized wind can remove angular
momentum from a thin disk rather efficiently.  Compared
to a viscous disk of the same sound-speed ($\cs$) and accretion
rate ($\dot M$), a wind-driven disk 
might have an accretion velocity that is larger by a factor 
$\sim r\Omega/\cs = r/h$.
The surface density would be correspondingly reduced, as would the
tendency toward self-gravity.

If viscous transport can be neglected, then under steady conditions,
\begin{equation}\label{vrwind}
\dot M_{\rm acc}\Omega r^2 = -\int\limits_{r_{\min}}^r
\overline{B_z B_\phi} r^{\prime\,2}\,dr'.
\end{equation}
The lefthand side is the angular-momentum flux carried inward
by the accreting gas.  The mass flux through the disk
$\dot M_{\rm acc}$, is now a function of $r$ even in a steady
disk because of the mass loss through the wind,
$\dot M_{\rm wind}\ll\dot M_{\rm acc}$.
The righthand side of eq.~(\ref{vrwind}) is the rate of loss
of angular momentum by maxwell stresses exerted on both faces
of the disk, assuming that $B_\phi$ is odd in $z$ while
$B_z$ is even.  The overbars denote an average over fluctuations
in azimuth and time.  The minus sign appears because we
define $\dot M_{\rm acc}>0$ for inflow.  The corresponding
equation for a viscous disk is
\begin{equation}\label{vrvisc}
\dot M_{\rm acc}\Omega r^2 = -\frac{1}{2}r^2\int\limits_{-h}^h
\overline{B_r B_\phi} \,dz.
\end{equation}
The magnetic part
of the angular-momentum flux emerges through 
surface areas $4\upi(r^2-r_{\min}^2)$ and $4\upi r h$
in eqs.~(\ref{vrwind}) \& (\ref{vrvisc}), respectively.
Since the former area is larger than the latter by a factor $r/h$,
the field needed to drive a given $\dot M_{\rm acc}$
is smaller by $h/r$ in a wind-driven disk than in a viscous disk.

If the effective viscosity of an
accretion disk is magnetic, then the energy density of the field
is at most in equipartition with the gas, \emph{viz}
$\beta_{\rm mag} \equiv 8\upi p_{\rm gas}/\mathbf{B}^2 \gtrsim 1$,
because
%a stronger field would quickly escape from the disk
%\emph{via} the Parker instability, and more importantly, because
a super-equipartion field would shut off the crucial magnetorotational
instability \citep{Balbus_Hawley98}.  We now give two arguments to show that
even in a wind-driven accretion disk,
the field must also be at or below equipartion.

Vertical hydrostatic equilibrium requires
\begin{displaymath}
\left.\frac{B_r^2+B_\phi^2-B_z^2}{8\upi}~+p_{\rm gas}\right|_{z=0}^{z=h}
\approx \int\limits_0^h \rho\Omega^2 z\,dz\,.
\end{displaymath}
We assume that the field has
dipolar symmetry, so that $B_z$ is
approximately constant with $z$ on the scale $h\ll r$, whereas
the horizontal components vanish at the midplane.
Since the righthand side above is positive, and since $p_{\rm gas}(z=h)\ll
p_{\rm gas}(z=0)$, it follows that
\begin{equation}\label{Blimit}
\frac{B_r^2+B_\phi^2}{8\upi}(z=\pm h) \lesssim p_{\rm gas}(z=0).
\end{equation}
Henceforth $B_r$ and $B_\phi$ are evaluated at $z=h$, and $p_{\rm \rm
gas}$ at $z=0$.  As pointed out by \cite{Blandford_Payne82}, in order that
centrifugal force should drive the wind outward, the poloidal field
lines must make an angle of at most $60^\circ$ with the surface of the
disk, so that $B_r/B_z\ge \sqrt{3}$.  Therefore
\begin{displaymath}
\frac{\mathbf{B}^2}{8\upi}\le \frac{4B_r^2+B_\phi^2}{8\upi}\le 4p_{\rm gas}.
\end{displaymath}
The magnetic energy density is less important to us, however, than the
azimuthal force per unit area on the disk:
\begin{equation}\label{maxforce}
\left|\frac{B_z B_\phi}{4\upi}\right|\le \sqrt{3}
\left|\frac{B_r B_\phi}{4\upi}\right| \le \sqrt{3}
\frac{B_r^2+B_\phi^2}{8\upi} < \sqrt{3}\,p_{\rm gas}.
\end{equation}
Using the inequality in eq.~(\ref{vrwind}), together with
$\dot M_{\rm acc}\equiv -2\upi\Sigma v_r$ and
$p_{\rm gas}\approx \Sigma\cs^2/2h$, we have
\begin{equation}\label{vrwind_est}
|v_r| \lesssim \frac{\sqrt{3}}{\Omega r^3\Sigma}
\int\limits_{r_{\min}}^r\frac{\Sigma\cs^2}{h}(r')\,r^{\prime\,2}\,dr'
\sim \sqrt{3}\frac{\cs^2}{\Omega h}.
\end{equation}
So the inflow could be marginally supersonic.
For a viscous disk, on the other hand, $|v_r|\approx\alpha\cs^2/\Omega r$,
which is strongly subsonic as long as $\alpha\ll r/h$.

The angular momentum extracted from the face of the disk by the field
must be carried off by the wind. It is problematic whether a
strong wind can be launched from the disk \citep{Ogilvie_Livio01}, and whether
rapid wind-driven accretion is stable \citep{Cao_Spruit01}.  Apart from these
difficulties, consideration of the magnetic flux,
\begin{displaymath}
\Phi(r)\equiv \int\limits_{r_{\min}}^r B_z(r',0)\,2\upi r' dr',
\end{displaymath}
leads to important constraints on the field strength and accretion rate.
Presumably $\Phi(r)$ should not change secularly.   If $B_z$
is predominantly of one sign, then the increase of $|\Phi|$ by advection
must be balanced by diffusion of the lines through the inflowing gas.
The drift velocity of the lines
$v_{\rm drift}=-v_r\sim \eta_{\rm eff}/r$ if $B_z$ varies
on scales $\sim r$, where $\eta_{\rm eff}$ is the effective diffusivity
of the gas.  In significantly ionized disks, the microscopic diffusivity
is negligible, so $\eta_{\rm eff}$ is due to turbulence, and one
expects $\eta_{\rm eff}=\alpha_{\rm mag}\cs h$ with
$\alpha_{\rm mag}\lesssim 1$ for much the same reasons that the
effective viscosity $\nu_{\rm eff}=\alpha_{\rm mag}\cs h$ with
$\alpha_{\rm visc}\lesssim 1$.  Hence for a steady wind-driven disk threaded
by net magnetic flux, $|v_r|\lesssim(\alpha_{\rm mag}h/r)\cs $, which
is probably very much less than the upper limit (\ref{vrwind_est}) and
comparable to the accretion velocity of a viscous disk.
Alternatively, the net flux could be essentially zero if $B_z$ changes
sign on scales $\ll r$.  In the latter case, the higher flow speed
(\ref{vrwind_est}) may be achievable.  But so irregular a field
would probably have to be sustained by dynamo action within the disk rather
than inherited from whatever region supplies the accreting gas.  This
probably requires magnetorotational instability (henceforth MRI)
and gives another argument for a sub-equipartion field.

To summarize this subsection, accretion driven by magnetized winds
is even less well understood than viscous accretion but might
allow substantially higher accretion velocities and lower surface
densities, perhaps by factors up to $\sim r/\alpha h$.

\subsection{Thin disks with strongly magnetized coronae}\label{corona}

This is a variant of \S\ref{windy} in which most of the field lines
are not open but re-attach to the disk at large
distances $\Delta r\gg h$ \citep{Galeev_Rosner_Vaiana79,Heyvaerts_Priest89}.
The vertical magnetic scale height is then
$H\sim\Delta r\gg h$.
The angular momentum flux carried through the corona,
\begin{displaymath}
\dot J_{\rm cor}(r)= -2\int\limits_{z=h}^{z=\infty}dr
\int\limits_0^{2\upi}d\phi\,\frac{B_r B_\phi}{4\upi}
~\sim~H\,\frac{B_r B_\phi}{4\upi}(r,h)\,,
\end{displaymath}
can be larger than the flux within the gas layer
by a factor $\sim H/h$, so that the effective value
of $\alpha$ might be as large as $r/h$ (for $H\sim r$) without
fields exceeding equipartion.

The evidence for magnetized coronae that dominate angular-momentum
transport is suggestive but inconclusive.  Local simulations of MRI
generally find that the scale height of the field exceeds that of the
gas, but only by factors of order unity; they also find $\alpha\sim
10^{-2}-10^{-1}$ rather than $\sim r/h$
\citep{Brandenburg_etal95,Stone_etal96,Miller_Stone00}.  Possibly, $H/h$ is
limited by the fact that the smallest dimension of the computational
domain is $\lesssim h$.  Global simulations of MRI have been performed
for relatively thick disks only, so that it is difficult to
distinguish scalings with $h$ from scalings with $r$
\citep{Matsumoto_Shibata97,Hawley00}.  Global simulations of
\emph{thin} disks in three dimensions may not be available for some
time because of the very large numbers of grid cells needed to resolve
both the disk and the corona.  \cite{Merloni_Fabian01} argue that
X-ray observations of accreting black holes (AGN and X-ray binaries)
demand a strongly magnetized corona, at least in the innermost part of
the disk.  On the other hand, observations of eclipsing cataclysmic
variables indicate that X-rays are emitted from the disk-star boundary
layer rather than an extended corona \citep{Mukai_etal97, Ramsay_etal01}.

\subsection{Global spiral waves}\label{spirals}

As is well known, a trailing $m$-armed spiral density wave
\begin{equation}\label{spiral}
\Sigma(r,\theta)= \Sigma_0(r) 
~+\Sigma_m(r)\cos\left(m\theta+\mu\ln r\right)
\end{equation}
exerts an outward (positive) gravitational angular-momentum flux 
\begin{displaymath}
\Gamma\approx \upi^2 G\Sigma_m^2r^3\,
\frac{m\mu}{|\mu|^{3}}
\end{displaymath}
\citep{Lynden-Bell_Kalnajs72}.
The above approximation is good for tightly-wrapped
waves, $\mu\gg m\ne0$, which carry relatively little flux
for a given density contrast $\Sigma_m/\Sigma_0$.  An exact
formula for logarithmic spirals
with $\Sigma_m\propto r^{-3/2}$ is
\begin{equation}\label{spiraltorque}
\Gamma = -\upi^2 G\Sigma_m^2r^3\,m\frac{\partial}{\partial\mu} K(\mu,m),
\end{equation}
where $K(\mu,m)$ is the Kalnajs function \citep{Kalnajs71}:
\begin{equation}\label{Kalnajs}
K(\mu,m)= \frac{1}{2}\left|\frac
{\Gamma\left[\frac{1}{2}(m+i\mu\,+\frac{1}{2})\right]}
{\Gamma\left[\frac{1}{2}(m+i\mu\,+\frac{3}{2})\right]}\right|^2\qquad
\mbox{(real $\mu$)}.
\end{equation}
With $\Sigma_m/\Sigma_0=1$, the largest ratio for which the
surface density is everywhere positive, one finds that the torque
is maximized at $m=1$ and a pitch angle $\tan^{-1}(m/\mu)\approx
48^\circ{\kern-4.5pt.2}$,
so that
\begin{equation}\label{maxtorque}
\Gamma_{\max}\approx 0.961\upi G\Sigma_0^2 r^3.
\end{equation}

Suppose that the gravitational torque is balanced by the
advection of angular momentum with the accreting gas.
In other words, 
$\Gamma \approx \dot M\Omega r^2$, 
so that there is no secular change in the angular momentum
within radius $r$.  The gravitationally-driven accretion speed is then
\begin{equation}\label{vrgrav}
|v_r|\le \frac{\Gamma_{\rm max}}{2\upi r^3\Omega\Sigma_0}
\approx 0.15 Q^{-1}\,\cs\,
\end{equation}
in which $Q$ has been calculated from the azimuthal average
of the surface density, $\Sigma_0$.  In principle therefore, accretion
may occur at a significant fraction of the sound speed.
But the existence of a selfconsistent wave-driven flow has
been assumed rather than proved.
Nonlinear single-armed spirals in keplerian disks have been
found by \cite{Lee_Goodman99}, but only for weak self-gravity
($Q\gg1$) and without dissipation or accretion.

In steady accretion onto a central mass that dominates the rotation
curve, the advected angular-momentum flux $\Gamma=\dot M\Omega
r^2\propto r^{1/2}$, so $\Sigma\propto r^{-5/4}$ rather than
$r^{-3/2}$.  Presumably the slight change in power-law index does not
change the results (\ref{maxtorque})-(\ref{vrgrav}) much.

\subsection{Clumpy disks}\label{clumpy}

Can QSO disks persist at $Q\ll1$ without fragmenting entirely into stars?
This question is all the more urgent because of rather
direct evidence for parsec-scale accretion disks in nearby AGN, if not 
QSOs, from VLBI observations of maser emission.
If the nuclear disk of NGC 1068 is in a steady state, then the
nuclear luminosity implies $Q\sim 10^{-3}$ at $r\sim 1\pc$ \citep{Kumar99}.
NGC 4258 is much less luminous, and estimates of $\dot M$ range
from $7\times 10^{-5}\alpha_{0.1}\dotsm$ based on
modeling the maser emission itself \citep{Neufeld_Maloney95},
to $10^{-2}\dotsm$ \citep{Gammie_Narayan_Blandford99,Kumar99} for
an assumed central ADAF; at the former rate, $Q\sim 1$ at the
outer edge of the masering region, $\sim 0.2\pc$ \citep{Maoz95},
while in the latter, $Q\sim 10^{-2}$.

\citet{Kumar99} has suggested a clumpy rather than smooth
disk, in which accretion occurs by gravitational scattering and
physical collisions among clumps rather than an $\alpha$ viscosity.
These clumps are supposed to be gas clouds rather than fully formed
stars, in order to provide appropriate conditions for maser amplification.
Although the model deals with the stability and accretion rate of the
clumpy disk as a whole, it does not ask
what prevents the individual clumps from collapsing.
In the application to NGC 1068, the masses, radii, and surface temperatures
of the clumps are quoted as $M_{\rm c}\sim 10^3M_\odot$, $R_{\rm c}\sim
0.1\pc$, and $T_{\rm eff,c}\approx 500\K$;
the virial temperature implied by this mass and radius is $\sim 2000\K$.
These clumps would be moderately optically thick at the wavelength
corresponding to $T_{\rm eff,c}$.  Their characteristic thermal 
time---the time required to radiate their binding energy---is
\begin{displaymath}
t_{\rm th,c}\approx\frac{GM_{\rm c}^2/R_{\rm c}}{4\upi R_{\rm c}^2
\sigma T_{\rm eff,c}^4}\sim  10^5\s\,,
\end{displaymath}
hence many orders of magnitude less than the orbital time
($\Omega^{-1}\sim 10^{11}\s$ at $r\sim 1\pc$) or interclump collision time.
Of course the surface temperatures of these objects are assumed to
be maintained by irradiation from the central source, but without
a fast-responding feedback mechanism, the thermal equilibrium is unstable:
if a clump starts to contract, its surface temperature will rise,
and collapse will proceed on the clump's internal free-fall
timescale (since this is larger than $t_{\rm th,c}$).

To answer the question raised above, we therefore believe that no QSO
accretion disk, whether smooth or clumpy, can persist at $Q\ll1$.

\section{Discussion}\label{discussion}

Although the accretion velocity of a thin viscous disk
is very strongly subsonic,  with a Mach number
$\Mach\equiv v_r/\cs\approx\alpha h/r$, the discussion of
\S\ref{alternatives} suggests that $\Mach$
might approach $\sim 0.1$ if accretion is driven by
large-scale magnetic or gravitational fields rather than a local
effective viscosity.  Self-consistent solutions that achieve
this bound would be very interesting to pursue.

Even at near-sonic accretion speeds,
QSO disks become selfgravitating at radii less than a parsec
if there is no important source of heating other than dissipation
of orbital energy.
Combining eqs.~(\ref{Teff}) \& (\ref{Tmid0})
with $\dot M=2\upi r\Sigma \Mach\cs$ instead
of the viscous relation (\ref{alphadef}), and assuming that
gas pressure dominates ($\beta\approx 1$), one finds that
\begin{eqnarray}\label{Qmach}
Q&=& 2^{-25/18}\upi^{-1}\Mach^{\,7/9}\krel^{2/9}\kes^{7/9}G^{-1}
\left(\frac{c\kB}{\sigma^{1/4} m}\right)^{8/9}\rs^{-11/9}
\left(\frac{r}{\rs}\right)^{-25/18}\nonumber\\
&\approx& 0.8\,\left(\frac{r}{10^4\rs}\right)^{-25/18}
\left(\frac{\Mach_{0.1}^7\veps_{0.1}^5\krel^2}{\lum^5 M_8^{11}}
\right)^{1/9}\nonumber\\
&\approx& 0.8\left(\frac{r}{0.1\pc}\right)^{-25/18}
\left(\frac{\Mach_{0.1}^7\veps_{0.1}^5\krel^2}{\lum^5}\right)^{1/9}
M_8^{1/6}\,.
\end{eqnarray}

On the other hand, we have seen  (\S\ref{energetics}) that
when angular-momentum transport is viscous, then
it is unlikely that stars can supply enough additional heat
to stabilize the disk beyond one parsec, especially if viscosity is
proportional to gas pressure rather than total pressure as
viscous stability probably demands.
In the latter case, even low-mass black holes embedded in the disk
are probably inadequate.  These statements
assume that enough free gas remains in the disk to supply the 
central black hole at its Eddington rate.  If all the gas converts
to stars, stability may result, but the quasar is quenched.
Perhaps a combination of stellar (or embedded-black-hole) heating
and a super-viscous accretion speed may allow an extended
gravitationally stable disk; we hope to explore this possibility
in a future paper.

Given the serious theoretical difficulties of all proposed mechanisms
for keeping $Q\gtrsim1$ at large radii, we are forced to take
seriously the only remaining possibility: that QSO disks do not exist
much beyond $\rsg\sim 10^{-2} \pc$ [eq.~(\ref{rsgrad})]---at least not
in a state of centrifugal support, vertical hydrostatic equilibrium,
and steady accretion.  Yet the mass within this radius is smaller than
that of the central black hole by a factor $\sim(h/r)\sim 10^{-2}$
(since the midplane density $\approx M/2\upi r^3$ at $Q=1$).  Hence in
order to grow the black hole, the disk must be replenished, either
steadily or intermittently, by infall of low-angular-momentum
material.  In order that the gas not circularize outside
$10^{-2}\rpc$, its specific angular momentum must be $\lesssim
70\,M_8^{1/2}\,\kms\,\pc$, or some three orders of magnitude
smaller than that of most stars in ellipticals and bulges
\citep{Binney_Merrifield98}.  Such a small ratio may seem unlikely, but
on the other hand, $M_{\rm bh}/M_{\rm bulge}\approx 10^{-3}$
\citep{McLure_Dunlop01}.  So perhaps the QSO is fueled by 
the low-angular-momentum tail of gas that forms the
bulge.  This gas would arrive at the outer edge of the disk in a
vertically broad infall, perhaps already carrying dust formed
from metals injected by outflows from the bulge or QSO disk itself,
and hence taking the place of a warped outer disk as the source
of reprocessed infrared light.
The picture this calls to mind is similar, except in scale,
to the standard scenario for the formation of a protostar
\citep{Shu_Adams_Lizano87}.

I thank Charles Gammie, Pawan Kumar, Kristen Menou, Ramesh Narayan,
and Bohdan Paczy\'nski for helpful discussions.  This work was
supported by the NASA Origins Program under grant NAG 5-8385.

\bibliographystyle{mn2e}

\begin{thebibliography}{}

\bibitem[\protect\citeauthoryear{{Ayal}, {Livio} \& {Piran}}{{Ayal}
  et~al.}{2000}]{Ayal_Livio_Piran00}
{Ayal} S.,  {Livio} M.,    {Piran} T.,  2000, ApJ, 545, 772

\bibitem[\protect\citeauthoryear{{Balbus} \& {Hawley}}{{Balbus} \&
  {Hawley}}{1998}]{Balbus_Hawley98}
{Balbus} S.~A.,  {Hawley} J.~F.,  1998, Rev. Mod. Phys., 70, 1

\bibitem[\protect\citeauthoryear{{Balbus} \& {Papaloizou}}{{Balbus} \&
  {Papaloizou}}{1999}]{Balbus_Papaloizou99}
{Balbus} S.~A.,  {Papaloizou} J.~C.~B.,  1999, ApJ, 521, 650

\bibitem[\protect\citeauthoryear{Binney \& Merrifield}{Binney \&
  Merrifield}{1998}]{Binney_Merrifield98}
Binney J.,  Merrifield M.,  1998, Galactic astronomy.
Princeton, NJ Princeton University Press

\bibitem[\protect\citeauthoryear{{Blandford} \& {Payne}}{{Blandford} \&
  {Payne}}{1982}]{Blandford_Payne82}
{Blandford} R.~D.,  {Payne} D.~G.,  1982, MNRAS, 199, 883

\bibitem[\protect\citeauthoryear{{Brandenburg}, {Nordlund}, {Stein} \&
  {Torkelsson}}{{Brandenburg} et~al.}{1995}]{Brandenburg_etal95}
{Brandenburg} A.,  {Nordlund} A.,  {Stein} R.~F.,    {Torkelsson} U.,  1995,
  ApJ, 446, 741+

\bibitem[\protect\citeauthoryear{{Cameron}}{{Cameron}}{1978}]{Cameron78}
{Cameron} A.~G.~W.,  1978, in Protostars and Planets: Studies of Star Formation
  and of the Origin of the Solar System, pp 453--487

\bibitem[\protect\citeauthoryear{{Cannizzo}, {Lee} \& {Goodman}}{{Cannizzo}
  et~al.}{1990}]{Cannizzo_Lee_Goodman90}
{Cannizzo} J.~K.,  {Lee} H.~M.,    {Goodman} J.,  1990, ApJ, 351, 38

\bibitem[\protect\citeauthoryear{Cao \& Spruit}{Cao \&
  Spruit}{2001}]{Cao_Spruit01}
Cao X.,  Spruit H.,  2001, astro-ph/0108484

\bibitem[\protect\citeauthoryear{{Collin} \& {Zahn}}{{Collin} \&
  {Zahn}}{1999a}]{Collin_Zahn99a}
{Collin} S.,  {Zahn} J.,  1999a, A\&A, 344, 433

\bibitem[\protect\citeauthoryear{{Collin} \& {Zahn}}{{Collin} \&
  {Zahn}}{1999b}]{Collin_Zahn99b}
{Collin} S.,  {Zahn} J.,  1999b, Ap\&SS, 265, 501

\bibitem[\protect\citeauthoryear{{Ebisuzaki}, {Makino}, {Tsuru}, {Funato},
  {Portegies Zwart}, {Hut}, {McMillan}, {Matsushita}, {Matsumoto} \&
  {Kawabe}}{{Ebisuzaki} et~al.}{2001}]{Ebisuzaki_etal01}
{Ebisuzaki} T.,  {Makino} J.,  {Tsuru} T.~G.,  {Funato} Y.,  {Portegies Zwart}
  S.,  {Hut} P.,  {McMillan} S.,  {Matsushita} S.,  {Matsumoto} H.,    {Kawabe}
  R.,  2001, ApJ, 562, L19

\bibitem[\protect\citeauthoryear{{Evans} \& {Kochanek}}{{Evans} \&
  {Kochanek}}{1989}]{Evans_Kochanek89}
{Evans} C.~R.,  {Kochanek} C.~S.,  1989, ApJ, 346, L13

\bibitem[\protect\citeauthoryear{{Galeev}, {Rosner} \& {Vaiana}}{{Galeev}
  et~al.}{1979}]{Galeev_Rosner_Vaiana79}
{Galeev} A.~A.,  {Rosner} R.,    {Vaiana} G.~S.,  1979, ApJ, 229, 318

\bibitem[\protect\citeauthoryear{{Gammie}}{{Gammie}}{2001}]{Gammie01}
{Gammie} C.~F.,  2001, ApJ, 553, 174

\bibitem[\protect\citeauthoryear{{Gammie}, {Narayan} \& {Blandford}}{{Gammie}
  et~al.}{1999}]{Gammie_Narayan_Blandford99}
{Gammie} C.~F.,  {Narayan} R.,    {Blandford} R.,  1999, ApJ, 516, 177

\bibitem[\protect\citeauthoryear{{Gebhardt}, {Bender}, {Bower}, {Dressler},
  {Faber}, {Filippenko}, {Green}, {Grillmair}, {Ho}, {Kormendy}, {Lauer},
  {Magorrian}, {Pinkney}, {Richstone} \& {Tremaine}}{{Gebhardt}
  et~al.}{2000}]{Gebhardt_etal00}
{Gebhardt} K.,  {Bender} R.,  {Bower} G.,  {Dressler} A.,  {Faber} S.~M.,
  {Filippenko} A.~V.,  {Green} R.,  {Grillmair} C.,  {Ho} L.~C.,  {Kormendy}
  J.,  {Lauer} T.~R.,  {Magorrian} J.,  {Pinkney} J.,  {Richstone} D.,
  {Tremaine} S.,  2000, ApJ, 539, L13

\bibitem[\protect\citeauthoryear{{Greenhill} \& {Gwinn}}{{Greenhill} \&
  {Gwinn}}{1997}]{Greenhill_Gwinn97}
{Greenhill} L.~J.,  {Gwinn} C.~R.,  1997, Ap\&SS, 248, 261

\bibitem[\protect\citeauthoryear{{Greenhill}, {Jiang}, {Moran}, {Reid}, {Lo} \&
  {Claussen}}{{Greenhill} et~al.}{1995}]{Greenhill_etal95}
{Greenhill} L.~J.,  {Jiang} D.~R.,  {Moran} J.~M.,  {Reid} M.~J.,  {Lo} K.~Y.,
    {Claussen} M.~J.,  1995, ApJ, 440, 619

\bibitem[\protect\citeauthoryear{{Hawley}}{{Hawley}}{2000}]{Hawley00}
{Hawley} J.~F.,  2000, ApJ, 528, 462

\bibitem[\protect\citeauthoryear{{Heyvaerts} \& {Priest}}{{Heyvaerts} \&
  {Priest}}{1989}]{Heyvaerts_Priest89}
{Heyvaerts} J.~F.,  {Priest} E.~R.,  1989, A\&A, 216, 230

\bibitem[\protect\citeauthoryear{{Hills}}{{Hills}}{1975}]{Hills75}
{Hills} J.~G.,  1975, Nature, 254, 295

\bibitem[\protect\citeauthoryear{{Ichimaru}}{{Ichimaru}}{1977}]{Ichimaru77}
{Ichimaru} S.,  1977, ApJ, 214, 840

\bibitem[\protect\citeauthoryear{{Illarionov} \& {Romanova}}{{Illarionov} \&
  {Romanova}}{1988}]{Illarionov_Romanova88}
{Illarionov} A.~F.,  {Romanova} M.~M.,  1988, AZh, 65, 682

\bibitem[\protect\citeauthoryear{{Kalnajs}}{{Kalnajs}}{1971}]{Kalnajs71}
{Kalnajs} A.~J.,  1971, ApJ, 166, 275+

\bibitem[\protect\citeauthoryear{Keady \& Kilcrease}{Keady \&
  Kilcrease}{2000}]{Keady_Kilcrease00}
Keady J.~J.,  Kilcrease D.~P.,  2000, in Cox A.~N.,  ed., , Allen's
  Astrophysical Quantities, 4 edn, AIP Press; Springer, New York, Chapt.~5, pp
  95--120

\bibitem[\protect\citeauthoryear{{Kumar}}{{Kumar}}{1999}]{Kumar99}
{Kumar} P.,  1999, ApJ, 519, 599

\bibitem[\protect\citeauthoryear{{Kurucz}}{{Kurucz}}{1992}]{Kurucz92}
{Kurucz} R.~L.,  1992, Revista Mexicana de Astronomia y Astrofisica, 23, 181

\bibitem[\protect\citeauthoryear{{Lacy}, {Townes} \& {Hollenbach}}{{Lacy}
  et~al.}{1982}]{Lacy_Townes_Hollenbach82}
{Lacy} J.~H.,  {Townes} C.~H.,    {Hollenbach} D.~J.,  1982, ApJ, 262, 120

\bibitem[\protect\citeauthoryear{{Lee} \& {Goodman}}{{Lee} \&
  {Goodman}}{1999}]{Lee_Goodman99}
{Lee} E.,  {Goodman} J.,  1999, MNRAS, 308, 984

\bibitem[\protect\citeauthoryear{{Lightman} \& {Eardley}}{{Lightman} \&
  {Eardley}}{1974}]{Lightman_Eardley74}
{Lightman} A.~P.,  {Eardley} D.~M.,  1974, ApJ, 187, L1

\bibitem[\protect\citeauthoryear{{Lynden-Bell} \& {Kalnajs}}{{Lynden-Bell} \&
  {Kalnajs}}{1972}]{Lynden-Bell_Kalnajs72}
{Lynden-Bell} D.,  {Kalnajs} A.~J.,  1972, MNRAS, 157, 1+

\bibitem[\protect\citeauthoryear{McLure \& Dunlop}{McLure \&
  Dunlop}{2001}]{McLure_Dunlop01}
McLure R.~J.,  Dunlop J.~S.,  2001, astro-ph/0108417

\bibitem[\protect\citeauthoryear{{Maoz}}{{Maoz}}{1995}]{Maoz95}
{Maoz} E.,  1995, ApJ, 455, L131

\bibitem[\protect\citeauthoryear{{Martin} \& {Kennicutt}}{{Martin} \&
  {Kennicutt}}{2001}]{Martin_Kennicutt01}
{Martin} C.~L.,  {Kennicutt} R.~C.,  2001, ApJ, 555, 301

\bibitem[\protect\citeauthoryear{{Matsumoto} \& {Shibata}}{{Matsumoto} \&
  {Shibata}}{1997}]{Matsumoto_Shibata97}
{Matsumoto} R.,  {Shibata} K.,  1997, in ASP Conf. Ser. 121: IAU Colloq. 163:
  Accretion Phenomena and Related Outflows, pp~443+

\bibitem[\protect\citeauthoryear{{McMillan}, {Lightman} \& {Cohn}}{{McMillan}
  et~al.}{1981}]{McMillan_Lightman_Cohn81}
{McMillan} S.~L.~W.,  {Lightman} A.~P.,    {Cohn} H.,  1981, ApJ, 251, 436

\bibitem[\protect\citeauthoryear{{Menou} \& {Quataert}}{{Menou} \&
  {Quataert}}{2001}]{Menou_Quataert01}
{Menou} K.,  {Quataert} E.,  2001, ApJ, 552, 204

\bibitem[\protect\citeauthoryear{Merloni \& Fabian}{Merloni \&
  Fabian}{2001}]{Merloni_Fabian01}
Merloni A.,  Fabian A.~C.,  2001, MNRAS, 321, 549

\bibitem[\protect\citeauthoryear{{Merritt} \& {Ferrarese}}{{Merritt} \&
  {Ferrarese}}{2001}]{Merritt_Ferrarese01}
{Merritt} D.,  {Ferrarese} L.,  2001, ApJ, 547, 140

\bibitem[\protect\citeauthoryear{{Miller} \& {Stone}}{{Miller} \&
  {Stone}}{2000}]{Miller_Stone00}
{Miller} K.~A.,  {Stone} J.~M.,  2000, ApJ, 534, 398

\bibitem[\protect\citeauthoryear{{Mukai}, {Wood}, {Naylor}, {Schlegel} \&
  {Swank}}{{Mukai} et~al.}{1997}]{Mukai_etal97}
{Mukai} K.,  {Wood} J.~H.,  {Naylor} T.,  {Schlegel} E.~M.,    {Swank} J.~H.,
  1997, ApJ, 475, 812+

\bibitem[\protect\citeauthoryear{{Nakai}, {Inoue} \& {Miyoshi}}{{Nakai}
  et~al.}{1993}]{Nakai_Inoue_Miyoshi93}
{Nakai} N.,  {Inoue} M.,    {Miyoshi} M.,  1993, Nature, 361, 45

\bibitem[\protect\citeauthoryear{{Narayan} \& {Yi}}{{Narayan} \&
  {Yi}}{1994}]{Narayan_Yi94}
{Narayan} R.,  {Yi} I.,  1994, ApJ, 428, L13

\bibitem[\protect\citeauthoryear{{Neufeld} \& {Maloney}}{{Neufeld} \&
  {Maloney}}{1995}]{Neufeld_Maloney95}
{Neufeld} D.~A.,  {Maloney} P.~R.,  1995, ApJ, 447, L17

\bibitem[\protect\citeauthoryear{{Ogilvie} \& {Livio}}{{Ogilvie} \&
  {Livio}}{2001}]{Ogilvie_Livio01}
{Ogilvie} G.~I.,  {Livio} M.,  2001, ApJ, 553, 158

\bibitem[\protect\citeauthoryear{{Paczy{\' n}ski}}{{Paczy{\'
  n}ski}}{1978}]{Paczynski78}
{Paczy{\' n}ski} B.,  1978, Acta Astron., 28, 91

\bibitem[\protect\citeauthoryear{{Ramsay}, {Poole}, {Mason}, {C{\' o}rdova},
  {Priedhorsky}, {Breeveld}, {Much}, {Osborne}, {Pandel}, {Potter}, {West} \&
  {Wheatley}}{{Ramsay} et~al.}{2001}]{Ramsay_etal01}
{Ramsay} G.,  {Poole} T.,  {Mason} K.,  {C{\' o}rdova} F.,  {Priedhorsky} W.,
  {Breeveld} A.,  {Much} R.,  {Osborne} J.,  {Pandel} D.,  {Potter} S.,  {West}
  J.,    {Wheatley} P.,  2001, A\&A, 365, L288

\bibitem[\protect\citeauthoryear{{Rees}}{{Rees}}{1978}]{Rees78}
{Rees} M.~J.,  1978, The Observatory, 98, 210

\bibitem[\protect\citeauthoryear{{Rees}, {Phinney}, {Begelman} \&
  {Blandford}}{{Rees} et~al.}{1982}]{Rees_Phinney_Begelman_Blandford82}
{Rees} M.~J.,  {Phinney} E.~S.,  {Begelman} M.~C.,    {Blandford} R.~D.,  1982,
  Nature, 295, 17

\bibitem[\protect\citeauthoryear{{Sanders}, {Phinney}, {Neugebauer}, {Soifer}
  \& {Matthews}}{{Sanders} et~al.}{1989}]{Sanders_etal89}
{Sanders} D.~B.,  {Phinney} E.~S.,  {Neugebauer} G.,  {Soifer} B.~T.,
  {Matthews} K.,  1989, ApJ, 347, 29

\bibitem[\protect\citeauthoryear{{Sellwood} \& {Balbus}}{{Sellwood} \&
  {Balbus}}{1999}]{Sellwood_Balbus99}
{Sellwood} J.~A.,  {Balbus} S.~A.,  1999, ApJ, 511, 660

\bibitem[\protect\citeauthoryear{{Shakura} \& {Sunyaev}}{{Shakura} \&
  {Sunyaev}}{1973}]{Shakura_Sunyaev73}
{Shakura} N.~I.,  {Sunyaev} R.~A.,  1973, A\&A, 24, 337

\bibitem[\protect\citeauthoryear{{Shapiro} \& {Teukolsky}}{{Shapiro} \&
  {Teukolsky}}{1985}]{Shapiro_Teukolsky85}
{Shapiro} S.~L.,  {Teukolsky} S.~A.,  1985, ApJ, 292, L41

\bibitem[\protect\citeauthoryear{{Shlosman} \& {Begelman}}{{Shlosman} \&
  {Begelman}}{1987}]{Shlosman_Begelman87}
{Shlosman} I.,  {Begelman} M.~C.,  1987, Nature, 329, 810

\bibitem[\protect\citeauthoryear{{Shlosman} \& {Begelman}}{{Shlosman} \&
  {Begelman}}{1989}]{Shlosman_Begelman89}
{Shlosman} I.,  {Begelman} M.~C.,  1989, ApJ, 341, 685

\bibitem[\protect\citeauthoryear{{Shu}, {Adams} \& {Lizano}}{{Shu}
  et~al.}{1987}]{Shu_Adams_Lizano87}
{Shu} F.~H.,  {Adams} F.~C.,    {Lizano} S.,  1987, ARAA, 25, 23

\bibitem[\protect\citeauthoryear{Spitzer}{Spitzer}{1978}]{Spitzer78}
Spitzer L.,  1978, Physical processes in the interstellar medium.
New York : Wiley-Interscience

\bibitem[\protect\citeauthoryear{{Spitzer} \& {Saslaw}}{{Spitzer} \&
  {Saslaw}}{1966}]{Spitzer_Saslaw66}
{Spitzer} L.~J.,  {Saslaw} W.~C.,  1966, ApJ, 143, 400+

\bibitem[\protect\citeauthoryear{{Stone}, {Hawley}, {Gammie} \&
  {Balbus}}{{Stone} et~al.}{1996}]{Stone_etal96}
{Stone} J.~M.,  {Hawley} J.~F.,  {Gammie} C.~F.,    {Balbus} S.~A.,  1996, ApJ,
  463, 656+

\bibitem[\protect\citeauthoryear{{Zel'Dovich} \& {Podurets}}{{Zel'Dovich} \&
  {Podurets}}{1966}]{Zeldovich_Podurets66}
{Zel'Dovich} Y.~B.,  {Podurets} M.~A.,  1966, Soviet Astronomy, 9, 742+

\end{thebibliography}

\begin{appendix}
\section{Viscously heated disks}
For completeness, this appendix gives formulae for the
midplane temperature ($T$), the surface density ($\Sigma$), and the
gravitational stability parameter ($Q$) in a steady disk heated
by viscous dissipation only, and cooled by radiative diffusion.

Combining eqs.~(\ref{alphadef}), (\ref{Teff}), and (\ref{Tmid0}),
and writing $\beta\cs^2=\kB T/m$, where $m\approx m_H$ is the mean
mass per gas particle, we have the radial dependence of $\Sigma$ \& $T$:
\ba\label{Tmid}
T&=& \left(\frac{\kappa m}{16\upi^2\alpha\beta^{b-1}\kB\sigma}\right)^{1/5}
\dot M^{2/5}\Omega^{3/5}\\
&\approx& 1.0\times10^5\left(\frac{\lum^2\krel}{\veps_{0.1}^2\alpha_{0.01}
\beta^{b-1}}\right)^{1/5} M_8^{-1/5}\left(\frac{10^3\rs}{r}\right)^{9/10}
\,\K,\nonumber
\ea
\ba\label{Sigmar}
\Sigma&=& \frac{2^{4/5}}{3\upi^{3/5}}\left(m^4\sigma\over\kB^4\right)^{1/5}
(\alpha\beta^{{b}-1})^{-4/5} \kappa^{-1/5}\dot M^{3/5}\Omega^{2/5}\\
&\approx& 3.9\times 10^6\,(\alpha_{0.01}\beta^{b-1})^{-4/5}
\lum^{3/5}\veps_{0.1}^{-3/5}\krel^{-1/5} M_8^{1/5}
\left(\frac{10^3\rs}{r}\right)^{3/5}~\mbox{g cm}^{-2}\,.
\nonumber
\ea
If viscosity scales with gas pressure (${b}=1$) then 
eqs.~(\ref{Sigmar})-(\ref{Tmid}) do not depend on $\beta$, which in
any case is not an independent parameter:
\bd
\frac{\beta}{1-\beta}=\frac{p_{\rm gas}}{p_{\rm rad}}=
\frac{3c\kB}{4\sigma m}\,\frac{\rho}{T^3}= \frac{3c}{8\sigma}
\left(\kB\over m\right)^{1/2}\beta^{1/2}\frac{\Sigma\Omega}{T^{7/2}}\,;
\ed
this leads to
\begin{equation}\label{betaval}
\frac{\beta^{(1/2)+({b}-1)/10}}{1-\beta}
%2^{1/20}\alpha^{-1/10}(\veps/\lum)^{4/5}\krel^{-9/10}\kes^{-1/10}
%c\left(\frac{\kB}{m\sigma^{1/4}}\right)^{2/5}
%\left(\frac{r}{10^3\rs}\right)^{21/20}
%\nonumber\\[5pt]
\approx 0.44\,\alpha_{0.01}^{-1/10}\veps^{4/5}\lum^{-4/5}
\left(\frac{r}{10^3\rs}\right)^{21/20}.
\end{equation}
So the importance of gas pressure increases monotonically with
radius in a steady, viscously heated disk.

Using eq.~(\ref{Tmid}) to eliminate $\cs$ from eq.~(\ref{MQrel}),
\ba\label{Qgpd}
Q&=& \frac{3}{(4\upi)^{3/5}}\,\alpha^{7/10}\beta^{(7{b}-12)/10}
\left(\kB\over m\sigma^{1/4}\right)^{6/5}
G^{-1}\dot M^{-2/5}\Omega^{9/10}\\
&\approx& 8.1\times 10^{-2}\alpha_{0.01}^{7/10}\beta^{(7{b}-12)/10}
\left(\frac{\veps_{0.1}}{\lum}\right)^{2/5}\krel^{3/10}
M_8^{-13/10}\left(\frac{10^3\rs}{r}\right)^{27/20}~.\nonumber
\ea

\end{appendix}

\label{lastpage}

\end{document}